\documentclass[aps,pra,twocolumn,showpacs,floatfix,superscriptaddress]{revtex4-1}

\usepackage{amsmath}    % need for subequations
\usepackage{graphicx}   % need for figures
\usepackage{verbatim}   % useful for program listings
\usepackage{color}      % use if color is used in text
\usepackage{subfigure}  % use for side-by-side figures
\usepackage{hyperref}   % use for hypertext links, including those to external documents and URLs
\usepackage{bbold}
\usepackage{amssymb}

\newcommand{\emaxA}{\epsilon_{s}^{(a)}}
\newcommand{\emaxB}{\epsilon_{s}^{(b)}}

\newcommand{\tr}[1]{\mathrm{Tr}[#1]}
\newcommand{\e}[1]{\mathrm{e}^{#1}}

\newcommand{\E}{\mathcal{E}_\epsilon}
\newcommand{\Tr}{{\rm Tr}}
\newcommand{\mgap}{\bf}
\providecommand{\proj}[1]{|{#1}\rangle\langle{#1}|}

\begin{document}

\title{Quantum non-Gaussianity witnesses in the phase space}
\author{Catherine Hughes}
\affiliation{QOLS, Blackett Laboratory, Imperial College London, London SW7 2BW, UK}
\author{Marco G. Genoni}
\affiliation{Department of Physics \& Astronomy, University College London, 
Gower Street, London WC1E 6BT, United Kingdom}
\affiliation{QOLS, Blackett Laboratory, Imperial College London, London SW7 2BW, UK}
\author{Tommaso Tufarelli}
\affiliation{QOLS, Blackett Laboratory, Imperial College London, London SW7 2BW, UK}
\author{Matteo G. A. Paris}
\affiliation{Dipartimento di Fisica, Universit\`a degli Studi di Milano, I-20133 Milano, Italy}
\author{M. S. Kim}
\affiliation{QOLS, Blackett Laboratory, Imperial College London, London SW7 2BW, UK}
\affiliation{School of Computational Sciences, Korea Institute for Advanced Study, Hoegiro 87, Dongdaemun, Seoul 130-722, Korea}
\date{\today}

\begin{abstract}
We address detection of quantum non-Gaussian states, i.e. nonclassical
states that cannot be expressed as a convex mixture of Gaussian states,
and present a method to derive a new family of criteria based on generic
linear functionals. We then specialize this method to derive witnesses
based on $s$-parametrized quasiprobability functions, generalizing
previous criteria based on the Wigner function. In particular we discuss
in detail and analyse the properties of Husimi Q-function based 
witnesses and
prove that they are often more effective than previous criteria in
detecting quantum non-Gaussianity of various kinds of non-Gaussian
states evolving in a lossy channel.
\end{abstract}

\maketitle
%%%%%%%%%%%%%%%%%%%%%%%%%%%%%%%%%%%%%%%%
\section{Introduction}
\label{sec:intro}
%%%%%%%%%%%%%%%%%%%%%%%%%%%%%%%%%%%%%%%%
The classification of quantum states of the harmonic oscillator
according to classical/non-classical and Gaussian/non-Gaussian
paradigms has been an ongoing focus of research in quantum information for some time now. A
number of criteria for the detection of non-classicality have been
introduced, based on phase-space distributions \cite{Wigner, Glauber,
Sudarshan,Lee91, Lee92, Arvind97, Arvind98, Marchiolli01, Paris01,
Dodonov02, Kenfack04,Paavola11}, ordered moments \cite{Shchukin05, Kiesel08,
Vogel08}, and information-theoretic arguments \cite{Simon00, Duan00,
Marian02, Giorda10, Ferraro12, Gehrke12, Buono10}. A particular
attention has been devoted both in characterizing the set of states with
positive Wigner function \cite{Hudson, Werner,
Mandilara,mari12,veitch13} and in distinguishing Gaussian and
non-Gaussian states \cite{Genoni07, Genoni08, Genoni10}. The different
measures of non-Gaussianity proposed have been for example used to
characterize experimentally generated non-Gaussian states
\cite{MarcoPhAdd,TWBcond,ThermalCond}, but they could not discriminate
between states that can be written as mixtures of Gaussian states, and
the so-defined {\em quantum non-Gaussian states}. 
\par
From a physical point
of view this is a particularly important distinction, as quantum
non-Gaussian states can be only produced by means of highly non-linear
processes, while states belonging to the Gaussian convex hull can be
generated by means of Gaussian operations only and classical
randomization. In \cite{QNGRadim,QNGWigner} the first attempt to detect
quantum non-Gaussianity was pursued, by deriving witnesses based
respectively on photon-number probabilities and on the Wigner function.
The criterion \cite{QNGRadim} has been already used to detect quantum
non-Gaussian states produced in different experimental settings
\cite{Jez11,Jez12,predojevic12}. 
\par 
We here present a framework to
derive QNG witnesses based on generic linear functionals. We apply these
results to the case of $s$-parametrized quasiprobability distributions,
generalizing the criteria obtained in \cite{QNGWigner} for the Wigner
function, to the Husimi Q-function ($s=-1$) and in general to any
distribution characterized by a parameter $s<0$. 
\par
The paper is structured as follows: in Section
\ref{sec:qngintro} we present the problem by defining QNG, while
Section \ref{sec:convexity} illustrates how to derive bounds of linear
functionals on the Gaussian convex hull, along with their most important
properties. In Section \ref{sec:QNGCriteria} we present sufficient but not necessary QNG witnesses
based on quasiprobability distributions. In Section \ref{sec:C1} the
effectiveness of these criteria are investigated for Fock states,
photon-added coherent states, and photon-subtracted squeezed states,
focusing in particular on the performances corresponding to the
different quasiprobability distributions considered. In Section
\ref{sec:Error} we illustrate how the uncertainty on the measured
average photon number, propagates to the derived bounds, for different
values of the parameter $s$. Section \ref{sec:conclusions} concludes the
paper with final discussions and remarks.
%%%%%%%%%%%%%%%%%%%%%%%%%%%%%%%%%%%%%%%%
\section{Quantum non-Gaussianity}
\label{sec:qngintro}
%%%%%%%%%%%%%%%%%%%%%%%%%%%%%%%%%%%%%%%%
We begin by recalling the definition of the Gaussian convex hull
\begin{equation}\label{setG}
\mathcal{G} = \left\{\rho\in\mathcal{B(H)}\: | \: \rho=\int d\boldsymbol{\lambda} \: p(\boldsymbol{\lambda})|\psi_G(\boldsymbol{\lambda})\rangle\langle\psi_G(\boldsymbol{\lambda})|\right\},
\end{equation}
where $p(\boldsymbol{\lambda})$ can be an arbitrary probability distribution, $|\psi_G(\boldsymbol{\lambda})\rangle$ are pure Gaussian states and $\mathcal{B(H)}$ is the set of bounded operators. In general, all pure single-mode Gaussian states can be parametrized as $|\psi_G (\boldsymbol{\lambda})\rangle = D(\alpha)S(\xi)|0\rangle$ where $D(\alpha)$ and $S(\xi)$ are respectively the displacement and squeezing operators with the standard form presented in \cite{BarRad}, $|0\rangle$ is the vacuum state, $\alpha,\xi$ are arbitrary complex numbers and $\boldsymbol{\lambda}=\{\alpha,\xi\}$. The set $\mathcal{G}$ includes mixed Gaussian states as they can always be decomposed in the form \eqref{setG}, but also non-Gaussian states, that is states having a non-Gaussian Wigner function, as mixtures of coherent and squeezed states.

In line with Refs.~\cite{QNGRadim,QNGWigner}, \textit{a quantum state $\rho$ is defined quantum non-Gaussian iff $\rho\notin\mathcal{G}$.} To understand the importance of QNG in Quantum Optics, consider a simple example: given a single-mode field  initially prepared in the vacuum state, it is easy to verify that states belonging to $\cal G$ can be prepared by applying a combination of Gaussian operations and classical randomization. In contrast, the preparation of a quantum non-Gaussian state $\rho\notin\cal G$ starting from the vacuum field can only be achieved by means of some non-Gaussian operation, such as the application of a highly non-linear Hamiltonian (i.e. more than quadratic in the mode operators) or probabilistic non-Gaussian operations as photon addition/subtraction \cite{MSKTutorial}.
%%%%%%%%%%%%%%%%%%%%%%%%%%%%
\section{Bounding linear functionals on the Gaussian convex hull}\label{sec:convexity}
%%%%%%%%%%%%%%%%%%%%%%%%%%%%
Before proceeding to specialize our analysis to phase-space quasiprobability
distributions, it is worthwhile to discuss the general approach we shall take in order to witness QNG. Suppose that a single-oscillator quantum state $\varrho$ is the output of some experiment. Assume that the data of our experiment allows us to estimate a certain quantity $\Phi[\varrho]$, where $\Phi$ is a {\it linear} functional on the space of quantum states, and a bound $n$ on the average photon number, that is $\Tr{[\varrho \hat a^\dagger \hat a]}\leq n$. 
Remarkably, it may be possible to gain some information on the QNG character of the state $\varrho$, solely based on those two quantities. To see this, let us consider
\begin{equation}
{\cal G}_n\equiv\{\varrho_G\in {\cal G}|{\sf Tr}[\varrho_G\,\hat a^\dagger \hat a]\leq n\},
\end{equation}
which can be easily seen to be convex subsets of $\mathcal{G}$ for any $n\geq0$, and define the function
\begin{equation}
	B(n)\equiv\min_{\varrho_G\in{\cal G}_{n}}\Phi[\varrho_G].\label{abstract-bound}
\end{equation}
In other words, $B(n)$ is the lowest possible value that $\Phi[\varrho]$ could take compatible with the assumptions  (i) $\varrho\in\mathcal{G}$; (ii) $\Tr{[\varrho \hat a^\dagger \hat a]}\leq n$. Hence, if our state verifies (ii), but we find the quantity $\Phi[\varrho]$ to be {\it below} $B(n)$, we must conclude that $\varrho\notin\mathcal{G}$ (conversely, finding $\Phi[\varrho]\geq B(n)$ must be interpreted as an inconclusive result). 

A key step in this procedure is the calculation of the function $B(n)$ for a given $\Phi$. In general, this can be seen as a problem of linear optimization over an infinite-dimensional parameter space [see Eq.~\eqref{abstract-bound}]. Luckily, the optimization can be dramatically simplified by exploting the properties of $B(n)$. It turns out that it is sufficient to look for the (constrained) minimum of $\Phi$ amongst the set of {\it pure} Gaussian states, and that of {\it Rank-2} mixtures of Gaussian states. Therefore, for a fixed $\Phi$ and $n$, only a finite number of parameters needs to be optimized in order to find $B(n)$. While one may be able to derive these results by applying standard techniques of convex analysis, we find it worthwhile to present their proof in our context in Appendix \ref{genproof}.

%%%%%%%%%%%%%%%%%%%%%%%%%%%%%%%%%%%%%%%%
\section{Quantum non-Gaussianity witnesses in phase-space}
\label{sec:QNGCriteria}
%%%%%%%%%%%%%%%%%%%%%%%%%%%%%%%%%%%%%%%%
In the present work we show that that the structure of the states in Eq.~\eqref{setG} implies nontrivial constraints on their associated quasiprobability distributions. As a consequence, we will be able to certify QNG when those constraints are violated. We start by recalling the results of \cite{QNGWigner}, where it was shown that the Wigner function of any quantum state $\varrho$ belonging to the Gaussian convex hull satisfies the following inequality
\begin{equation}\label{W0}
W[\varrho](0) \geq \frac 2\pi \exp\{-2n(1+n)\}\:, \qquad n=\hbox{Tr}[\varrho a^\dag a] \:.
\end{equation}
We aim at obtaining bounds for other $s$-parametrized quasiprobabilities, which we express as a convolution \cite{convolution}. For a quantum state of density operator $\varrho$, 
\begin{equation}\label{Qs}
Q_s [\varrho] (\alpha) = \frac{1}{\pi^2}\int d^2\xi\ \chi[\varrho](\xi,s)\e{\alpha\xi^*-\alpha^*\xi}.
\end{equation} 
Here $\chi[\varrho](\xi,s)$ is the $s$-ordered characteristic function
\begin{equation}
\chi[\varrho](\xi,s) = \tr{\varrho\hat{D}(\xi)}\e{s|\xi|^2/2}.
\end{equation}
There are three values of $s$ for which the quasiprobability function is typically explored: $s=1$ is the Glauber-Sudarshan P-function \cite{Glauber, Sudarshan}, $s=0$ is the Wigner function \cite{Wigner}, and $s=-1$ is the Husimi Q-function \cite{Husimi}
For the purposes of this paper, the only necessary requirement on the parameter $s$ is going to be $s<0$, in order to avoid singularities in our quasiprobability distributions. Even though the function in Eq.~\eqref{Qs} may lose some of the appealing properties of a quasiprobability distribution when $s<-1$, it still allows us to obtain useful and experimentally friendly QNG criteria as we will discuss in Sec. \ref{sec:C1}.

\subsection{General QNG criteria in Phase space}\label{crit}
The general problem under investigation can be formulated in the general framework of Section~\ref{sec:convexity}, by noting that $Q_s[\varrho](\alpha)$ is a linear functional of the state at fixed $\alpha$ and $s$. Thus, having fixed a particular value of $s<0$, and assuming $\Tr[\varrho a^\dagger a]\leq n$, we ask ourselves whether the structure given in Eq.~\eqref{setG} implies a non-trivial lower bound on the possible values that $Q_s$ can take. More precisely, we define
\begin{equation}
{B}_s(n)\equiv \min_{\varrho\in{\cal G}_n}Q_s[\varrho](0). \label{mixbound}
\end{equation}
For every value of $s<0$, $B_s(n)$ is positive and convex. Moreover, we show in Appendix~\ref{quasi-general} that $B_s(n)$ is strictly decreasing in $n$, $B_s(n)\to0$ as $n\to\infty$, and the minimizing state in ${\cal G}_n$ has an average photon number exactly equal to $n$. The functions $B_s$ are therefore non-trivial and can be exploited in the formulation of QNG criteria as follows.

\textbf{Criterion 1:} For a quantum state $\varrho$, define the QNG witness
\begin{equation}
\Delta_s^{(a)}[\varrho]=Q_s[\varrho](0)-B_s(\bar{n}) \label{eq:witness1}
\end{equation}
where $\hbox{Tr}[\varrho a^\dag a]\leq \bar{n}$. Then,
\begin{equation}
\label{eq:Criterion1}
\Delta_s^{(a)}[\varrho]<0 \Longrightarrow\varrho\notin\mathcal{G},
\end{equation}
that is, $\varrho$ is quantum non-Gaussian.\\

\textbf{Criterion 2:} Consider now a quantum state $\varrho$ and a Gaussian map $\mathcal{E}_G$, or a convex mixture of such maps. Define:
\begin{equation}
\Delta_s^{(b)}[\varrho,\mathcal{E}_G]=Q_s[\mathcal{E}_G(\varrho)](0)-B_s(\bar{n}_{\mathcal{E}}) \label{eq:witness2}
\end{equation}
where $\hbox{Tr}[\mathcal{E}_G (\varrho) a^\dag a ]\leq \bar{n}_{\mathcal{E}}$.
Then,
\begin{equation}
\label{eq:Criterion2}
\exists\ \mathcal{E}_G\ \mbox{s.t.}\ \Delta_s^{(b)}[\varrho,\mathcal{E}_G]<0\Longrightarrow\varrho\notin\mathcal{G}.
\end{equation}
The proof of Eq. (\ref{eq:Criterion2}) is the same as that for Eq. (\ref{eq:Criterion1}) except that a Gaussian map $\mathcal{E}_G$ has now first been applied to the state. This results in a change in the mean photon number, but does not impact the procedure. A full proof is provided in \cite{QNGWigner}. Before proceeding further, we note that the monotonicity of $B_s$ implies that the criteria become harder to satisfy as $\bar n$ and $\bar{n}_{\mathcal{E}}$ are increased (indeed, both $\Delta_s^{(a)}$ and $\Delta_s^{(b)}$ would correspondingly increase). Therefore, in the remainder of this paper we shall apply these witnesses respectively for $\hbox{Tr}[\varrho a^\dag a]= \bar{n}$ and
$\hbox{Tr}[\mathcal{E}_G (\varrho) a^\dag a ]= \bar{n}_{\mathcal{E}}$, which provide the highest chance of detecting QNG. On the other hand, experimentally it may be more practical to estimate an upper bound to the average photon number, rather than its actual value. It is important to note that these criteria provide sufficient but not necessary bounds.

\subsection{Near-optimality of pure states}
As discussed in Section~\ref{sec:convexity}, we can restrict the optimization in Eq~\eqref{mixbound} to Rank-1 and Rank-2 mixtures of Gaussian states. In all the considered examples, however, we found strong numerical evidence that the minimum was being reached by a pure Gaussian state. We have thus proven the near-optimality of pure Gaussian states for a number of $s$-values of interest through a semi-analytical approach, whose details are provided in Appendix~\ref{pure-quasi}. A pure state lower bound to each quasiprobability can be defined as
\begin{equation}\label{purebound}
B^P_s(n)\equiv \min_{|\psi_G\rangle}\left\{Q_s[|\psi_G\rangle\langle\psi_G|](0)\,\,|\,\,\langle\psi_G| a^\dagger a|\psi_G\rangle\leq n \right\},
\end{equation}
where the $|\psi_G\rangle$'s are pure Gaussian states. Clearly, the bound in Eq.~\eqref{purebound} is in practice easier to calculate than the one in Eq.~\eqref{mixbound}, however $B^P_s(n)\geq B_s(n)$, since we can not exclude that the minimum may be reached by a Rank-2 state. Nevertheless, our numerical studies for the cases $s=\{ -1/4, -1/2, -1,-2,-3\}$ provide the bound
\begin{equation}\label{approx}
\left| B^P_s(n)-B_s(n)\right|\lesssim n\cdot 10^{-15},
\end{equation}
meaning that the pure state lower bound $B^P_s$ is an excellent approximation to the true bound $B_s(n)$ in a wide range of average photon numbers [see Appendix~\ref{pure-quasi}]. Direct calculations relating to this bound are shown in more detail in Appendix~\ref{purestatebounds}.
The level of approximation provided by Eq.~\eqref{approx} is sufficient to guarantee the validity of our findings in the following section.

%%%%%%%%%%%%%%%%%%%%%%%%%%%%%%%%%%%%%%%%
\section{Detecting quantum non-Gaussianity of states evolving in a lossy channel}
\label{sec:C1}
\begin{figure}[h!]
\includegraphics[width=0.95\columnwidth]{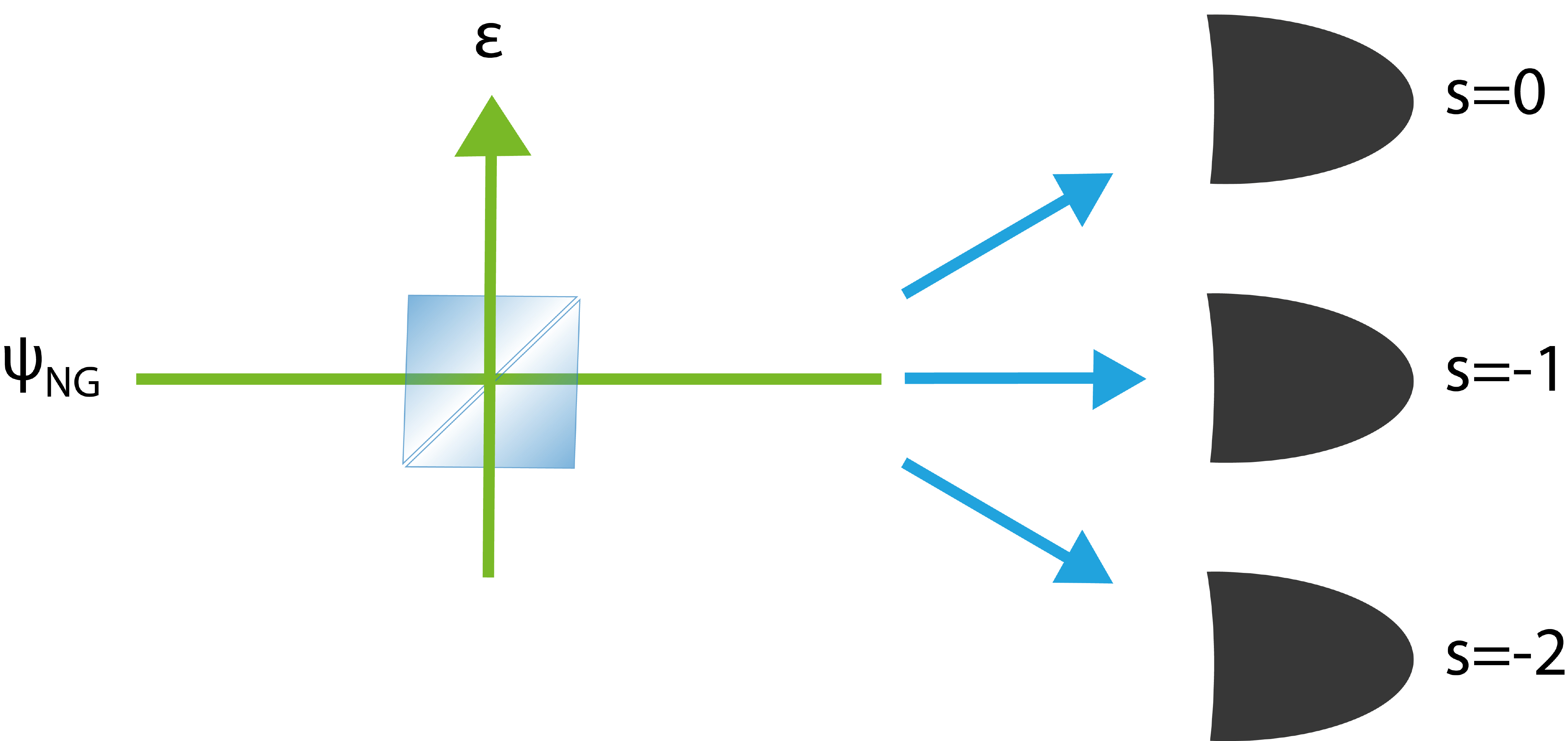}
\caption{We send a non-Gaussian state through a channel with loss $\epsilon$ and choose the detector with which to measure it. The $s=0$ detector correspond to parity measurement, $s=-1$ to the probability of vacuum detection, while $s=-2$ corresponds to an inefficient vacuum detection with efficiency $\eta=2/3$.
\label{f:Explanation}}
\end{figure}

%%%%%%%%%%%%%%%%%%%%%%%%%%%%%%%%%%%%%%%%
In this section we will test the effectiveness of the criteria introduced in Section~\ref{crit}. Specifically, we shall investigate whether these criteria can be exploited to certify that pure non-Gaussian states evolving in a lossy channel remain quantum non-Gaussian. 
We will consider initially pure non-Gaussian states evolving in a lossy bosonic channel described by the following master equation:
\begin{equation}
\dot{\varrho}=\frac{\gamma}{2}(\hat{a}\varrho\hat{a}^{\dagger}-\hat{a}^{\dagger}\hat{a}\varrho)+h.c.
\end{equation}
The corresponding quantum channel  $\E$ is Gaussian and can be characterized by a single parameter, $\epsilon=1-e^{-\gamma t}$. We will look for the maximum values of $\epsilon$ such that the criteria are violated, in particular we define
\begin{align}
\emaxA [\varrho] &= \max\{ \epsilon \: : \Delta_s^{(a)}[\E(\varrho)] \leq 0 \} \:, \\
\emaxB [\varrho] &= \max\{ \epsilon \: :\exists\mathcal{E}_{\sf G}\text{ s.t.} \ \Delta_s^{(b)}[\E(\varrho),\mathcal{E}_{\sf G}] \leq 0 \} \:.
\end{align}
Since for {\mgap $\epsilon>\frac12$}, no negativity of the Wigner function can be
observed, we will be interested in larger values of $\emaxA$ and
$\emaxB$, so that our criteria will be able to detect quantum
non-Gaussian states with positive Wigner function. The usefulness of the
Wigner-function-based criterion has been extensively shown in
\cite{QNGWigner,QNGCats}. We will start by comparing the witnesses
$\Delta_s^{(a)}$ for initial Fock states, while next we will discuss
both the witnesses $\Delta_s^{(a)}$ and $\Delta_s^{(b)}$ for initial
photon-added coherent states and photon-subtracted squeezed states. For
$\Delta_s^{(a)}$ we compare across three values of the parameter $s$:
the special cases $s=0$, $s= -1$ corresponding respectively to the
Wigner and Husimi-Q functions, and adding a third case at $s=-2$. The
quasiprobability distributions in the origin of phase-space, can be
evaluated as 
\begin{align}
Q_s[\varrho](0) = \frac{2}{\pi (1-s)}\sum_m (-1)^m 
\left(\frac{1+s}{1-s}\right)^m \langle m |\varrho|m\rangle \:.
\end{align}
It depends only on the photon-number probabilities $p_m =\langle m
|\varrho|m\rangle $, and thus can be in principle experimentally
measured by means of photon-number resolving detectors.  More in
particular the Wigner function in the origin corresponds to the average
value of the parity operator $\Pi=(-)^{a^\dag a}$, while for 
$s\rightarrow -1$, i.e. for the Husimi Q-function, we have 
the projection over the vacuum state 
$Q_{-1}[\varrho](0) = \langle 0|\varrho | 0\rangle$ which is measurable 
both by means of {\em on-off} and photon-number resolving
detectors or through heterodyne detection. If we rather consider values
of $s<-1$, it is possible to prove that $Q_s[\varrho](\alpha)$
corresponds to the rescaled heterodyne probability distribution,
obtained by means of detectors with efficiency $\eta=2/(1-s)$, such that
for $s=-2$ we have $\eta=2/3$ \cite{matteoQSM}. While the parameter $\epsilon$
characterizing the lossy channel is supposed to be unknown, and our goal
is to understand the maximum value of noise such that our criteria will
be able to detect quantum non-Gaussian states, the inefficiency of the
detector is known to the experimentalist as it is possible to determine
its value by probing the detector with known states. In fact, as
illustrated in Fig. \ref{f:Explanation}, considering different values of
$s$ is equivalent to detecting QNG of unknown states evolved through a
lossy channel, with a choice of detectors, one corresponding to each
$s$.
\par
As regards the examination of $\Delta_s^{(b)}$ , we will focus on
the special cases of $s=0$ and $s=-1$. In both cases we will observe
how, in particular in the low energy regime, the witnesses derived for
lower values of $s$ show a larger robustness against loss.

%

%%%%%%%%%%%%%%%%%%%%%%%%%%%%%%%%%%%%%%%%
\subsection{Fock states}
%%%%%%%%%%%%%%%%%%%%%%%%%%%%%%%%%%%%%%%%
A Fock state $|m\rangle$ evolves in a lossy channel as
\begin{align}
\E(|m\rangle\langle m|)=\sum^m_{l=0}{{m}\choose{l}}(1-\epsilon)^l\epsilon^{m-l}|l\rangle\langle l| \:,
\end{align}
and the value of the corresponding $s$-parametrized quasiprobability distribution at the origin can
be evaluated using the formula for a generic Fock state
\begin{align}
Q_s[|m\rangle\langle m |](0) = \frac{2}{\pi (1-s)} (-1)^m \left(\frac{1+s}{1-s}\right)^m \:.
\end{align}
\begin{figure}[h!]
\includegraphics[width=0.95\columnwidth]{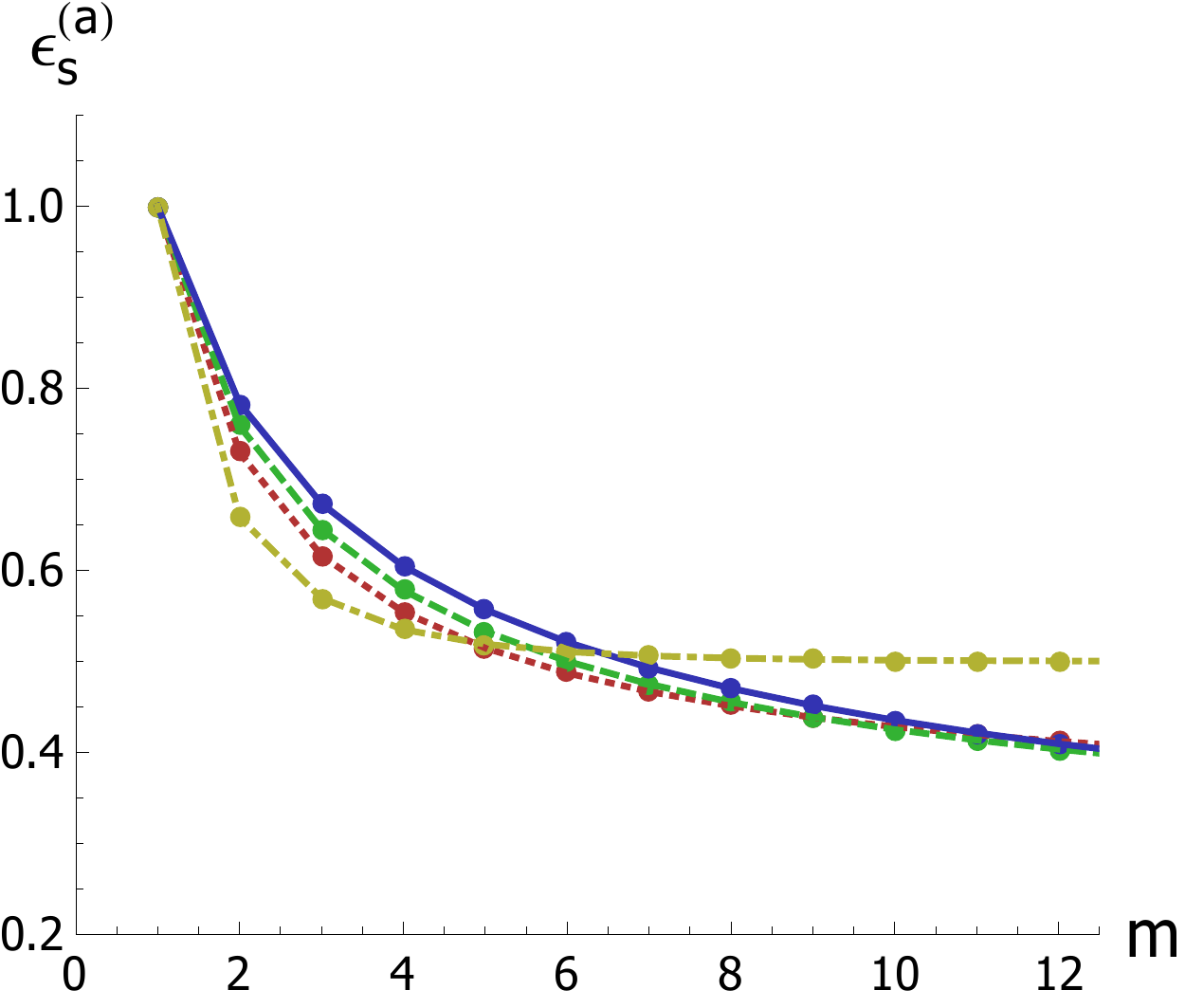}
\caption{Maximum value of the loss parameter $\emaxA$ such that the bounds 
are violated, as a function of the initial Fock number $m$ and for different values
of $s$: yellow (dot-dashed), $s=0$; red (dotted), $s=-1/2$; green (dashed), $s=-1$; blue (solid), $s=-2$.
\label{f:FockEpsilon}}
\end{figure}
We have evaluated the corresponding values of the witnesses $\Delta_s^{(a)}$, along
with the maximum values of the noise parameter $\emaxA$ where the bounds are
violated. In particular these are plotted in Fig. \ref{f:FockEpsilon} as a function of the 
Fock number $m$ and for different values of the $s$ parameter. Perhaps surprisingly, the witnesses appear to be more sensitive as $s$ decreases, that is, they provide a larger value of $\emaxA$ in the relevant range $m\leq 5$.

\begin{figure}[h!]
\hfill
\subfigure[\label{f:FockDelta1}]{\includegraphics[width=0.4\columnwidth]{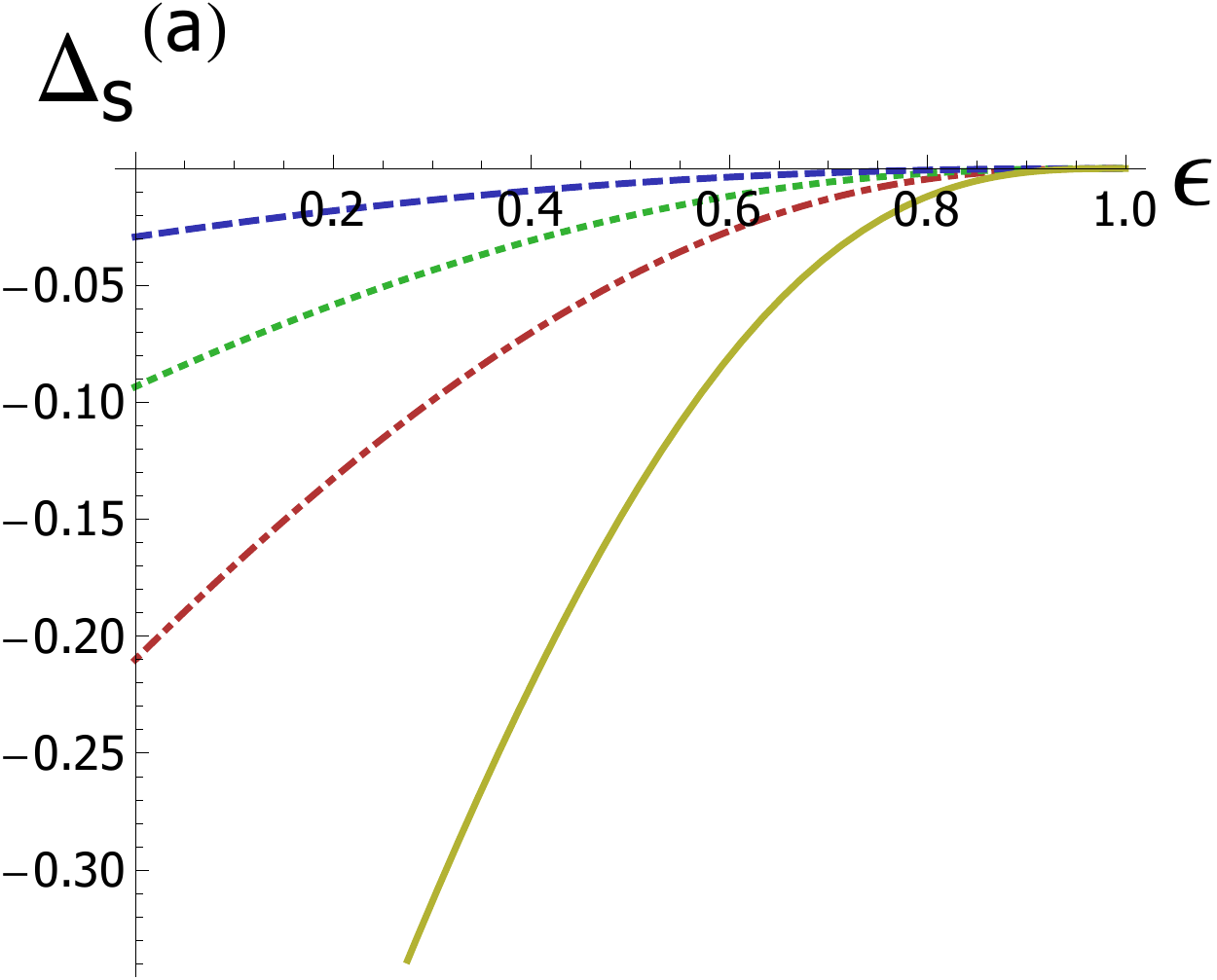}}
\hfill
\subfigure[\label{f:FockDelta3}]{\includegraphics[width=0.4\columnwidth]{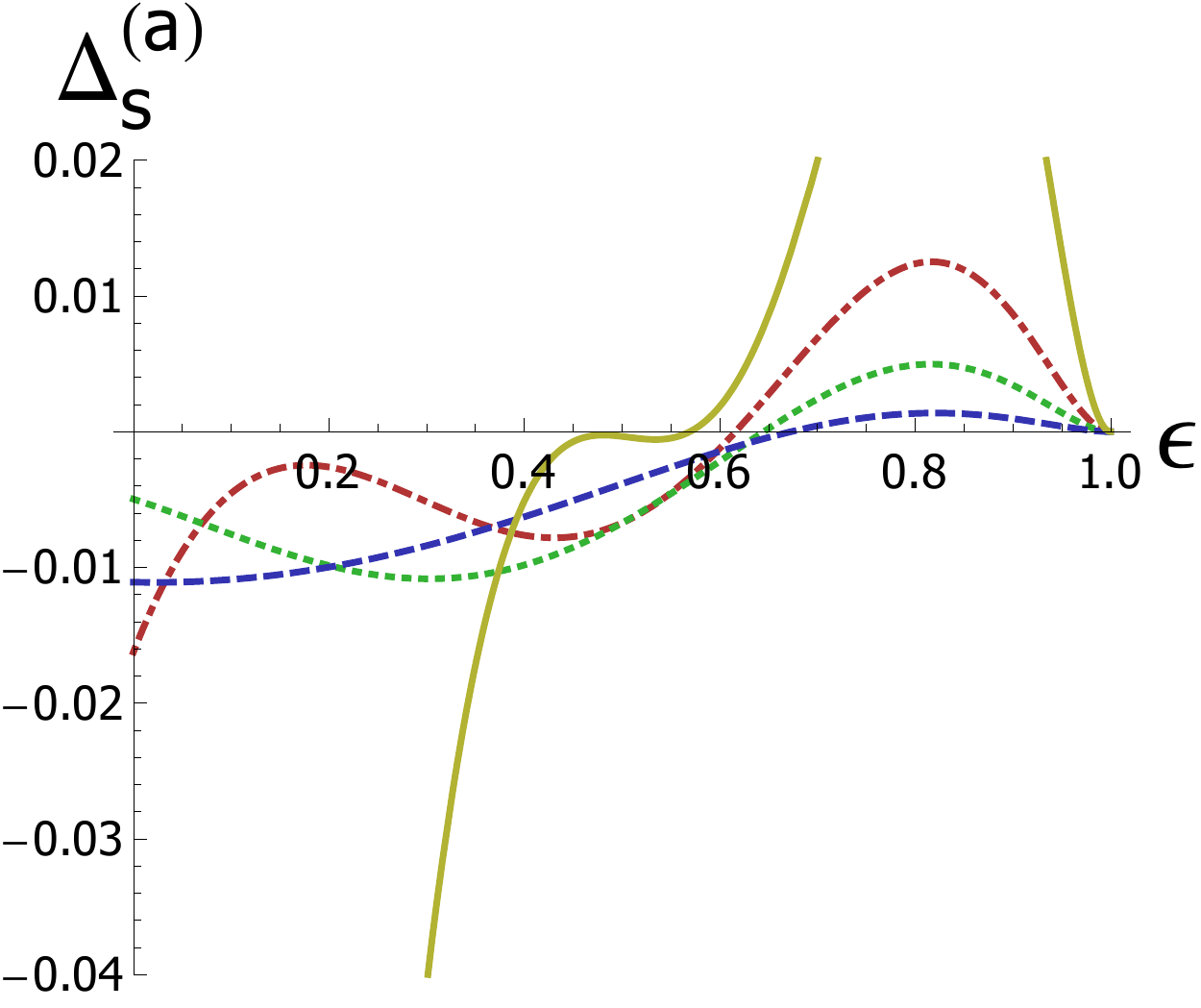}}
\hfill
\caption{QNG witnesses $\Delta_s^{(a)}$ for initial Fock states (a) $|1\rangle$ and (b) $|3\rangle$ as a function of the lossy parameter $\epsilon$ and for different values of $s$: yellow (solid), $s=0$; red (dot-dashed), $s=-1/2$; green (dotted), $s=-1$; blue (dashed), $s=-2$.}
\end{figure}

One can notice an interesting tradeoff in the behaviour of the witnesses $\Delta_s^{(a)}$ for the Fock state $|1\rangle$ in Fig. \ref{f:FockDelta1}. We note that the absolute value of $\Delta_s^{(a)}$ is decreasing by considering more negative values of $s$. Even though such monotonous behaviour is lost for the Fock state
$|3\rangle$ [see Fig. \ref{f:FockDelta3}], similar conclusions can be drawn for higher Fock states. Hence, while one can in principle detect QNG for larger values of the noise parameter by decreasing $s$, the amount of violation quantified by $\Delta_s^{(a)}$ may be generally smaller. The impact of such tradeoff on the experimental detection of QNG, however, cannot be assessed without a thorough analysis of the propagation of experimental errors for the various witnesses. A first attempt towards this direction will be done in Sec. \ref{sec:Error}, while a complete analysis goes beyond the scope of our paper.

%%%%%%%%%%%%%%%%%%%%%%%%%%%%%%%%%%%%%%%%
\subsection{PAC states}
%%%%%%%%%%%%%%%%%%%%%%%%%%%%%%%%%%%%%%%%
A photon-added coherent state is defined as
\begin{equation}
|\psi_{\tiny\mbox{PAC}}\rangle = \mathcal{N}\hat{a}^{\dagger}|\alpha\rangle,
\end{equation}
where $\mathcal{N}$ is the normalization factor. Its average photon number is
\begin{equation}
\bar{n}^{\tiny\mbox{PAC}}_0 = \frac{\alpha^4+3\alpha^2+1}{1+\alpha^2},
\end{equation}
for $\alpha\in\mathbb{R}$. The $s$-parametrized quasiprobability distributions are determined using the convolution expression presented in \cite{convolution}:
\begin{equation}
\label{eq:convolution}
Q_{s'}[\varrho](\alpha) = \frac{2}{\pi(s-s')}\int d^2\beta\  Q_s[\varrho](\beta)\ \e{-\frac{2|\alpha-\beta|^2}{(s-s')}},
\end{equation}
with the condition that this holds provided $s'<s$. Using this expression, the values of the witnesses $\Delta^{(a)}_s$ can be computed.
\begin{figure}[h!]
\includegraphics[width=0.95\columnwidth]{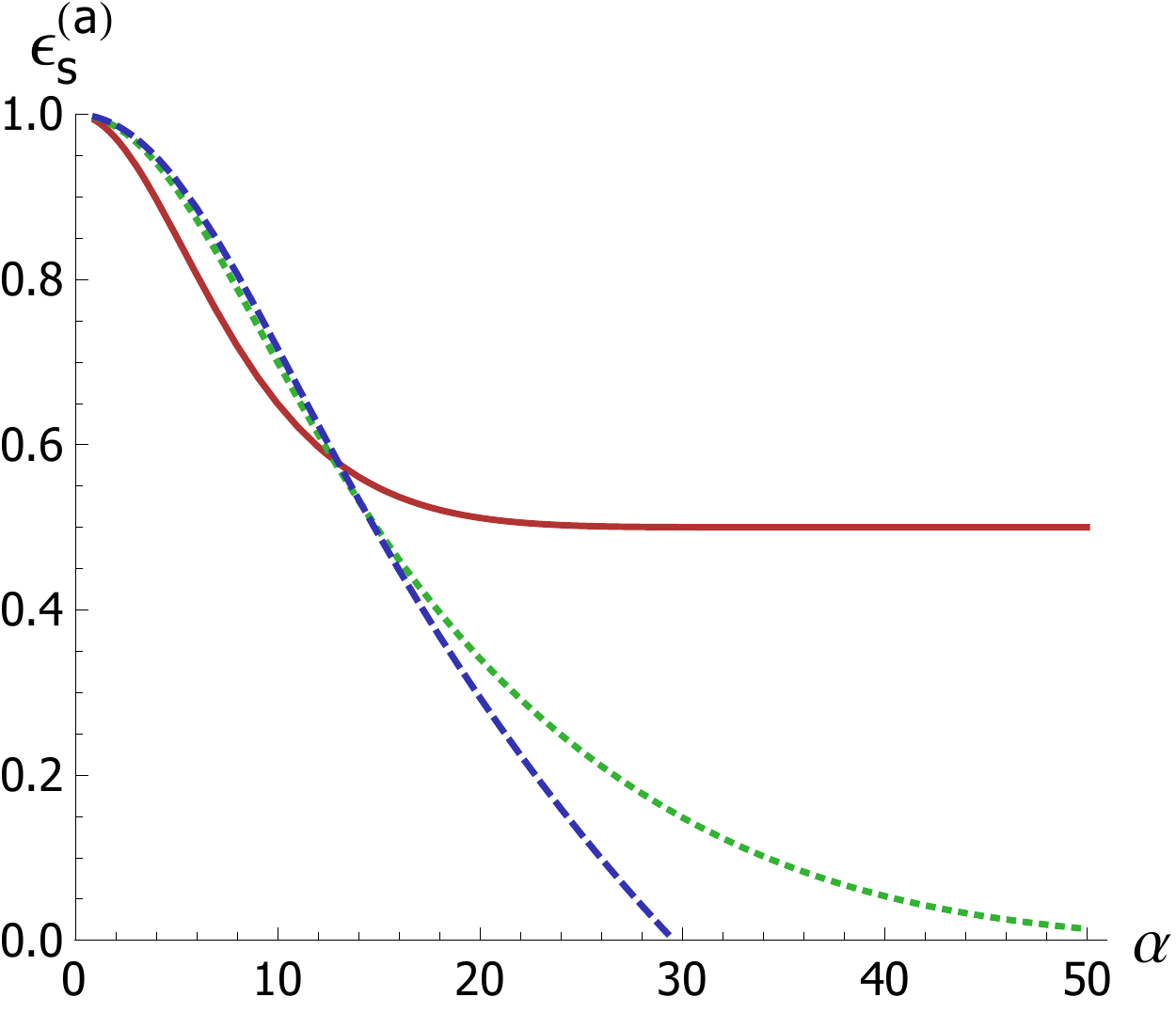}
\caption{Maximum value of the loss parameter $\emaxA$ such that the bounds 
are violated, as a function of the coherent state parameter $\alpha$ and for different values
of $s$: red (solid), $s=0$; green (dotted), $s=-1$; blue (dashed), $s=-2$.\\
\label{f:PACEpsilon}}
\end{figure}
Similarly to what was observed for the Fock states, smaller values of $s$ produce a more effective bound for the certification of QNG in noisy PAC, provided the parameter $\alpha$ is smaller or equal to about 10, as illustrated in Fig. \ref{f:PACEpsilon}. However, we observe again evidence that there is a compromise between a tighter bound and the amount of violation quantified by the criterion $\Delta_s^{(a)}$, with the magnitude of the parameter decreasing for lower values of $s$.

We can now consider the optimized witness defined in Eq. (\ref{eq:witness2}). For the PAC state, it is observed that the minima of the witness $\Delta^{(b)}_s[\varrho,D(\beta)]$ for the quasiprobability distributions are displaced from the origin of the phase space, and it is therefore possible to decrease the quantum non-Gaussianity indicator by re-displacing the minimum to the origin. Thus the quasiprobability function $Q_s[\varrho](-\beta)$ and the average photon number of the displaced state are computed, yielding:
\begin{equation}
\bar{n}^{\tiny\mbox{PAC}}(\beta)=(1-\epsilon)|\beta|^2\bar{n}^{\tiny\mbox{PAC}}_0+\sqrt{1-\epsilon}(\beta^*\langle\hat{a}\rangle_0+\beta\langle\hat{a}^{\dagger}\rangle_0),
\end{equation}
where $\langle A\rangle_0=\langle\psi_0|A|\psi_0\rangle$, and for $|\psi_0\rangle=|\psi_{\tiny\mbox{PAC}}\rangle$,
\begin{equation}
\langle\hat{a}\rangle_0=\langle\hat{a}^{\dagger}\rangle_0=\frac{\alpha(2+\alpha^2)}{1+\alpha^2}.
\end{equation}
We then minimize $\Delta^{(b)}_s[\varrho,D(\beta)]$ over the possible displacement parameters $\beta$. We find that the optimal value of $\beta$ for large values of $\epsilon$ and $\alpha\gneq  1.5$, which is nearly the same for the Wigner and Q functions, can be approximated as
\begin{equation}
\beta_{\tiny\mbox{opt}}\simeq-\alpha\sqrt{1-\epsilon}=-\alpha e^{-\gamma t/2}.
\end{equation}

Taking $\beta=\beta_{\tiny\mbox{opt}}$, we compare the values of the QNG witness based on the second criterion for $s=0$ and $s=-1$. Both the plots and numerical investigations indicate that $\epsilon_s^{(b)}\simeq 1$ for all possible values of $\alpha$. No root can be found for general $\alpha$.
%%%%%%%%%%%%%%%%%%%%%%%%%%%%%%%%%%%%%%%%
\subsection{PSS states}
%%%%%%%%%%%%%%%%%%%%%%%%%%%%%%%%%%%%%%%%

Taking the squeezing parameter $r$ to be real, define the photon-subtracted squeezed state as $|\psi_{\tiny\mbox{PSS}}\rangle=\mathcal{N}\hat{a}S(r)|0\rangle.$ The average photon number for the PSS state is
\begin{equation}
\bar{n}^{\tiny\mbox{PSS}}_0=3\sinh^2 r+1.
\end{equation}

\begin{figure}[h!]
\hfill
\subfigure[\label{f:PSSEpsilon}]{\includegraphics[width=0.45\columnwidth]{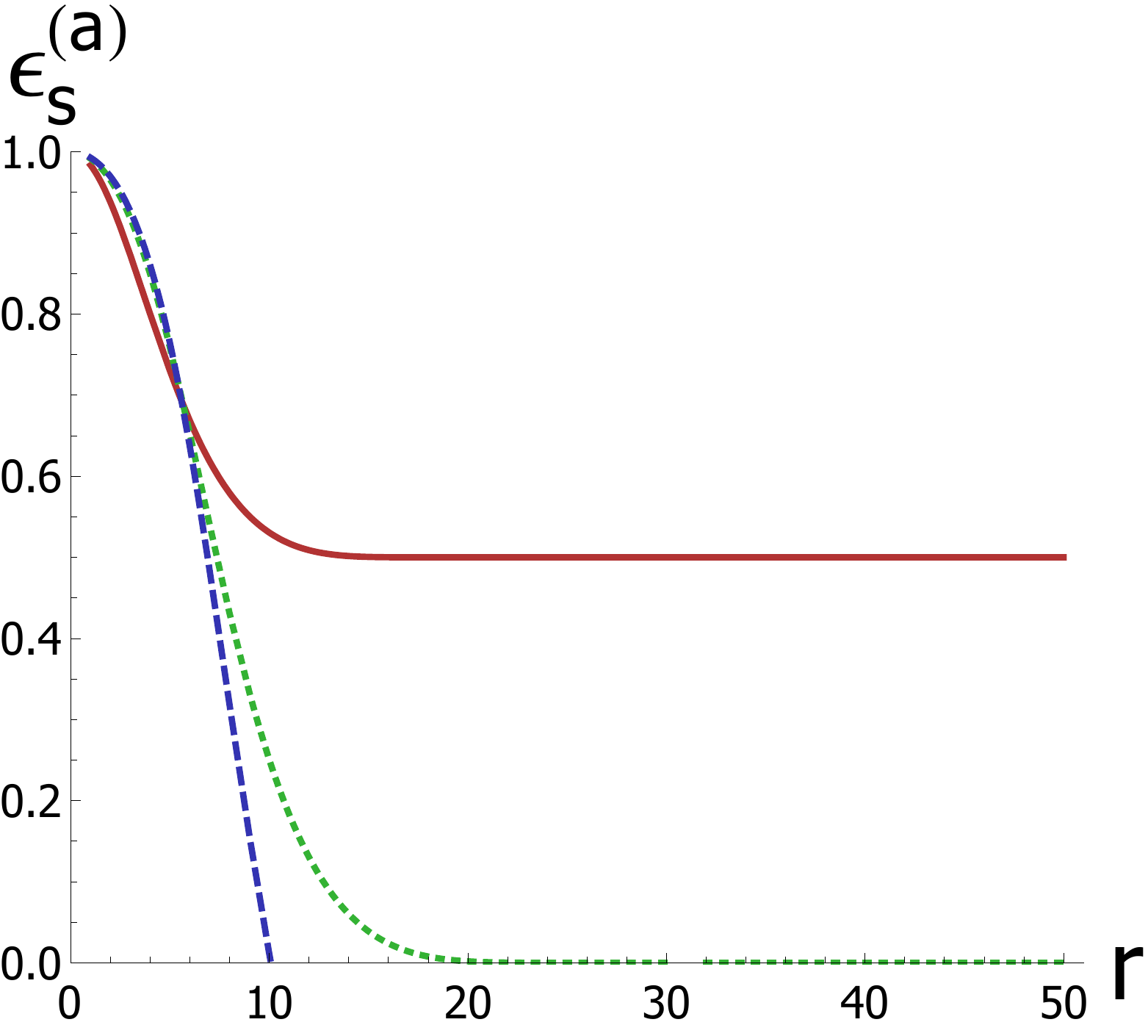}}
\hfill
\subfigure[\label{f:PSSEpsilonb}]{\includegraphics[width=0.45\columnwidth]{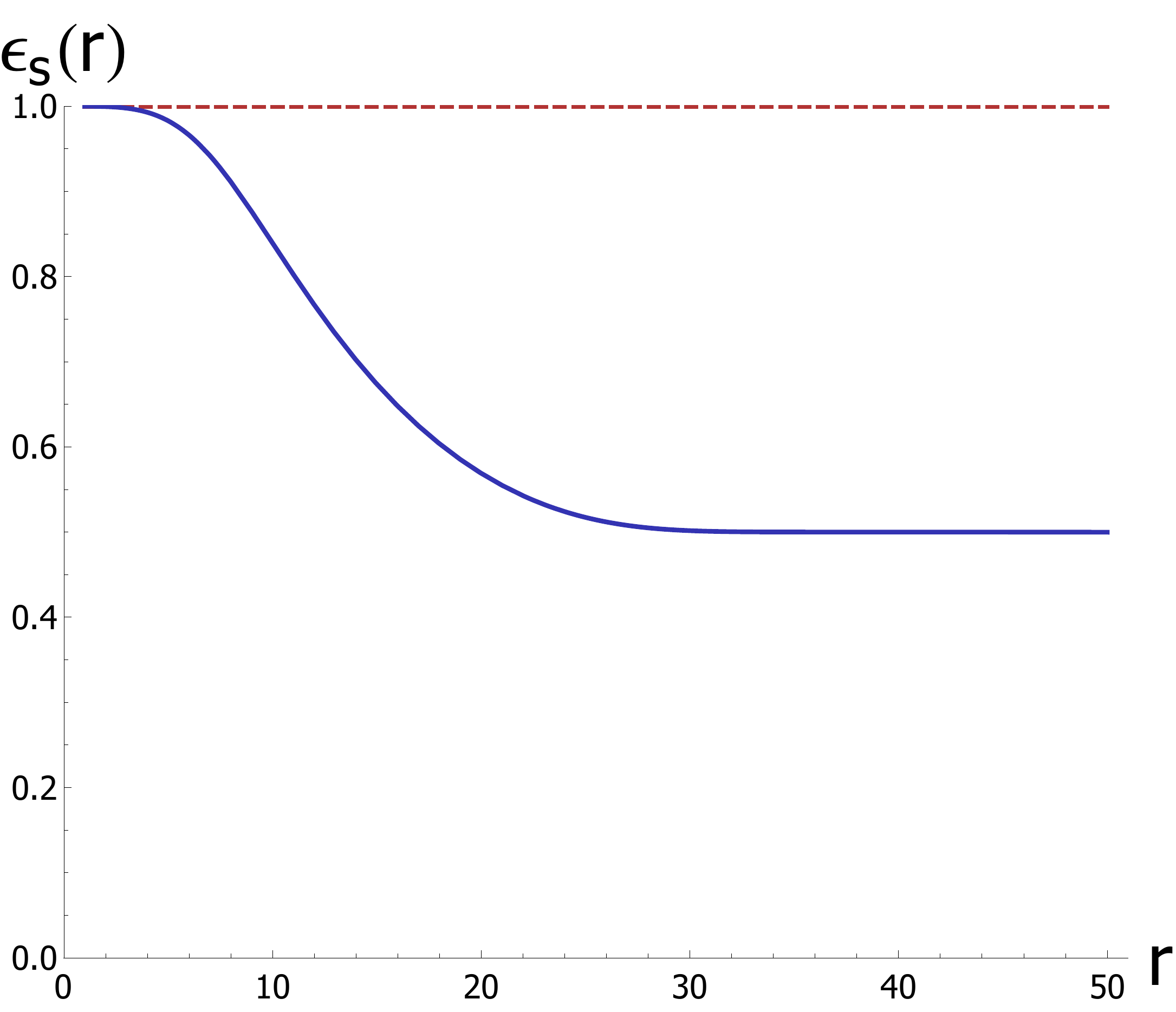}}
\hfill
\caption{Loss parameters for PSS states. (a) Maximum value of the loss parameter $\emaxA$ such that the bounds 
are violated, as a function of the squeezing parameter $r$ and for different values
of $s$: red (solid), $s=0$; green (dotted), $s=-1$; blue (dashed), $s=-2$. (b) A comparison between the maximum values of the noise parameter $\epsilon^{(b)}_s$ for PSS states by using the Wigner function criterion (blue solid line) and the Q function criterion (red dashed).}
\end{figure}

The quasiprobability distributions for the PSS state are computed using the convolution of Eq. \ref{eq:convolution}. The bounds were found, and again demonstrate the same characteristics found with the other states. In this case, more negative values of $s$ allow for a larger value of the loss parameter $\emaxA$ for squeezing parameter $r\lesssim 8$ [Fig. \ref{f:PSSEpsilon}]. Again, though, this represents a loss in the quantity of violation described by the parameter $\Delta_s^{(a)}$.

Similarly to how PAC states inherit a displacement, PSS states inherit additional squeezing on the evolved state, and we can use the optimized witness (\ref{eq:witness2}) by using additional squeezing operations. In this case, there is a difference between how the Wigner function changes under this squeezing versus how the Q function changes: while the Wigner function at the origin is unchanged by the squeezing operation, the Q function is not invariant and as a result the following argument for the optimization of the squeezing parameter is only valid for the Wigner function. This proves not to be a problem, and will be discussed below. First, we determine the value $q_{\tiny\mbox{opt}}$ that minimizes the average photon number of $S(q)\rho S^{\dagger}(q)$,
\begin{equation}
\bar{n}^{\tiny\mbox{PSS}}(q)=(1-\epsilon)[\bar{n}^{\tiny\mbox{PSS}}_0(\mu^2_q+\nu^2_q)+\mu_q\nu_q(\langle\hat{a}^2\rangle_0+\langle\hat{a}^{\dagger 2}\rangle_0)]+\nu^2_q,
\end{equation}
where $\mu_t=\cosh t,\ \nu_t=\sinh t$ and for an initial PSS state $|\psi_{\tiny\mbox{PSS}}\rangle$, $\langle\hat{a}^2\rangle_0 = \langle\hat{a}^{\dagger 2}\rangle_0 = 3\mu_r\nu_r.$
In this case the optimal squeezing value can be evaluated analytically:
\begin{align}
q_{\tiny\mbox{opt}} &= -\mbox{arccosh}(\mu_{\tiny\mbox{opt}}) \\
\mu_{\tiny\mbox{opt}} &= \frac{1}{\sqrt{2}}\left(1+\frac{6(1-\epsilon)\mu^2_r+4\epsilon-3}{\sqrt{(4\epsilon-3)^2+12(1-\epsilon)\epsilon\mu^2_r}}\right)^{1/2}.
\end{align}

We again follow the format of the PAC state analysis, assigning the squeezing parameter its optimal value $q_{\tiny\mbox{opt}}$ and plotting the criterion as a function of $\epsilon$. We plot both the Wigner and Q functions for $q_{\tiny\mbox{opt}}$ and illustrate that while the $q_{\tiny\mbox{opt}}$ used is only optimized for the Wigner function, the maximum noise $\emaxB$ for the Q function for this value of $s$ is 1 for all values and therefore already giving the desired result. So even if this is not the optimal squeezing for the Q function, it is sufficient to detect quantum non-Gaussianity by means of the Q function based witness. This feature is illustrated in Fig. \ref{f:PSSEpsilonb}.
%%%%%%%%%%%%%%%%%%%%%%%%%%%%%%%%%%%%%%%%
\section{Estimation of Error on the Bounds \label{sec:Error}}
\begin{figure}[h!]
\includegraphics[width=0.95\columnwidth]{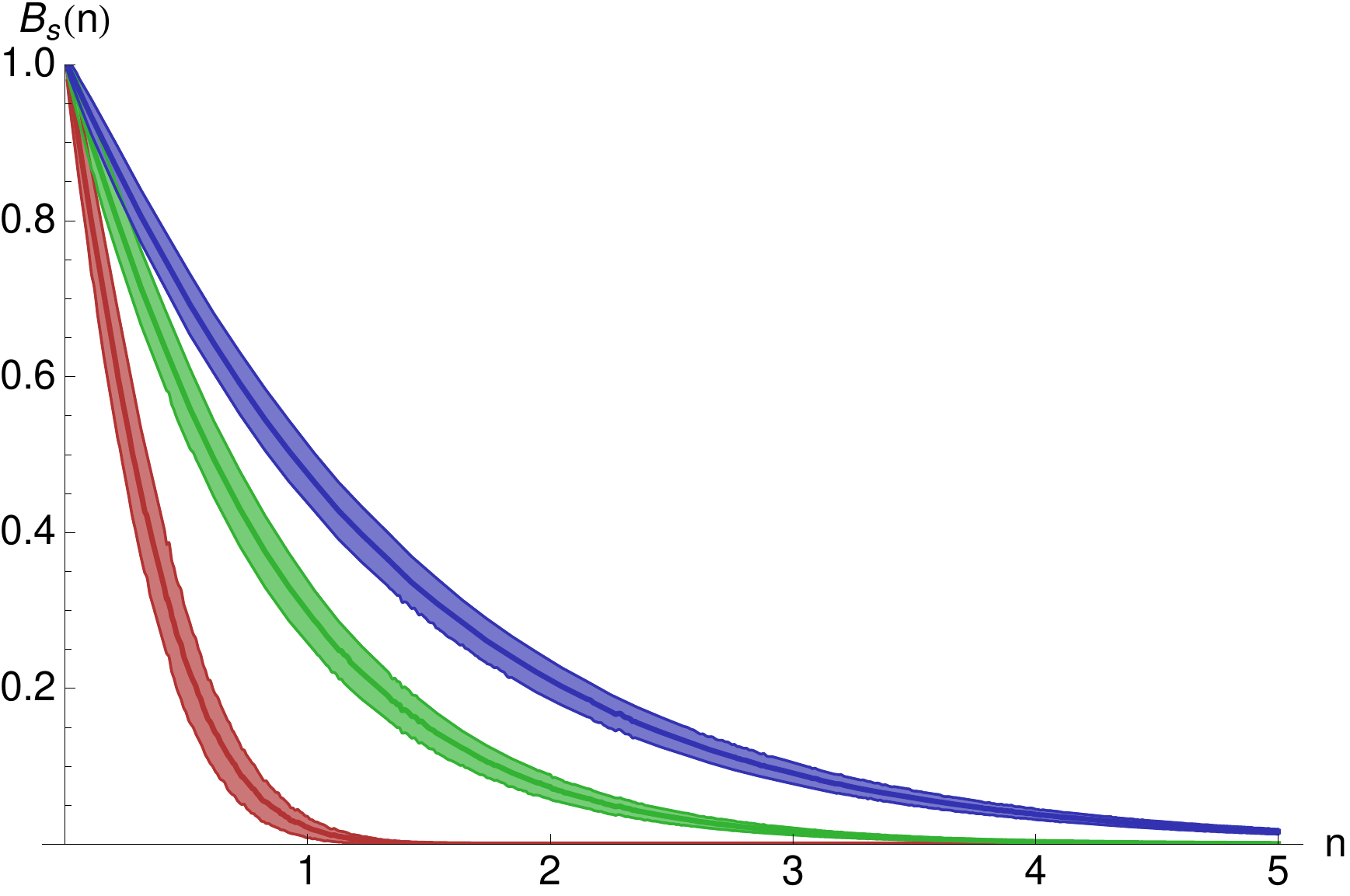}
\caption{Error on the (renormalized) bounding functions $B_s(n)$ for (from top to bottom) s=0 (blue), s=-1 (green), and s=-2 (red).\\
\label{f:ErrorBound}}
\end{figure}

%%%%%%%%%%%%%%%%%%%%%%%%%%%%%%%%%%%%%%%%
While a full error propagation to evaluate the various witnesses is beyond the scope of this paper, it is straightforward to evaluate the bounds $B_s(n)$ for the different $s$-values, based on uncertainty in the mean photon number $n$. The method of determining the error on the bound is chosen to best approximate the experimental realities. To this end, we suppose that we have a photon number resolving detector with which we would like to measure different average values of $n$ that we assign to a set $n_{avg}$. These values are a discretized version of the range of $n$ we would like to consider. In an experiment, we need to measure our state $k$ times, for preferably large $k$. We then define $n_{tot}$ as the total number of photons measured over all $k$ trials, that is:
\begin{equation}
n_{tot}=k\times n_{avg}.
\end{equation}
For $n_{tot}$ we assume a Poissonian distribution for simplicity. Before evaluating the means and variances of the bounds we divide by $k$ as we wish to evaluate for a distribution about $n_{avg}$. Normalizing the results so all bounds evaluate to 1 at $n_{avg}=0$, we get the distributions in Fig. \ref{f:ErrorBound}. As we can observe, the errors on the different bounds $B_s(n)$ are comparable. This shows how the errors coming from the measurement of the quasiprobability distributions values $Q_s[\varrho](\alpha)$, will probably play a major role when the proposed witnesses will be used in an actual experiment.
%
%%%%%%%%%%%%%%%%%%%%%%%%%%%%%%%%%%%%%%%%
\section{Conclusions} \label{sec:conclusions}
%%%%%%%%%%%%%%%%%%%%%%%%%%%%%%%%%%%%%%%%
We have presented a general method to derive bounds of linear functionals on the Gaussian convex hull. We note that these bounds are sufficient but not necessary for the characterization of this set. After having presented the main properties of the bounds, we used it to define QNG witnesses based on $s$-parametrized quasiprobability distributions, with $s<0$. The witnesses are based on bounding from above the average photon number of the quantum state, and measuring the value of the corresponding quasiprobability distribution in a particular point of phase space (typically the origin).

Following the determination of these witnesses, we consider three different states and test the criteria for three different values of $s$ for each state.  Motivation to consider the bound for different $s$-values comes from the freedom it provides to change the type of detection used in experiment. While it is known that $s=0,-1$ correspond to the Wigner and Q functions respectively, $s=-2$ is comparable to measuring the Q function with an inefficient detector. As the inefficiency of the detector can be known from trials using known states, allowing for $s<-1$ provides a more general description less dependent on the type of detection. From the different states for which the bound was considered we see that there is a region for which a smaller $s$-value provides a witness for QNG and allows more channel loss than the originally considered Wigner function bound. There is, however, a tradeoff between the maximum amount of loss for which QNG may be witnessed and the amount of violation quantified by the criterion which is generally smaller, for smaller values of $s$.

\vspace{-0.1cm}

\subsection*{Acknowledgements}
We thank M. Barbieri, D. Dorigoni, S. Olivares and R. Filip for fruitful discussions.
MGG acknowledge support from EPSRC (EP/K026267/1). TT and MSK
acknowledge support from the NPRP 4-554-1-084 from Qatar National
Research Fund. MGAP acknowledges support from MIUR (FIRB-LiCHIS-
RBFR10YQ3H). CH, MGG, TT, and MSK acknowledge support from EPSRC (EP/K034480/1).

\

\appendix
\section{General properties of the bounds}\label{genproof}
In this section we prove that if the minimum of $\Phi$ on ${\cal G}_n$ is not achieved by a pure Gaussian state, then it must be achieved by a Rank-2 mixture of pure Gaussian states. Before proving this, it is useful to introduce an auxiliary lemma.
\paragraph*{Lemma 1}
The function $B(n)$ is convex.
\paragraph*{Proof}
Let $\rho_1\in{\cal G}_{n_1}$ and $\rho_2\in{\cal G}_{n_2}$ be such that $B(n_1)=\Phi[\rho_1]$ and $B(n_2)=\Phi[\rho_2]$, and take $0\leq p\leq 1$. Then
\begin{align}
	&pB(n_1)+(1-p)B(n_2)=p \Phi[\rho_1]+(1-p)\Phi[\rho_2]\nonumber\\
	&=\Phi[p\rho_1+(1-p)\rho_2]\geq B(pn_1+(1-p)n_2),
\end{align}
where we have used the linearity of $\Phi$ and the fact that $p\rho_1+(1-p)\rho_2\in {\cal G}_{pn_1\!+\!(1\!-\!p)n_2}$ might not be the state which minimizes $\Phi$ in that set. 

\paragraph*{Theorem 1}
Given $n$, either there exists a pure Gaussian state $|\psi_{n}\rangle$ of ${\cal G}_{n}$ such that $B(n)=\Phi[|\psi_{n}\rangle\langle \psi_{n}|]$, or $B(n)=\Phi[\rho_n]$, where $\rho_n\in{\cal G}_n$ is a rank-2 state of the form $\rho_n=p|\psi_{1}\rangle\langle \psi_{1}|+(1-p)|\psi_{2}\rangle\langle \psi_{2}|$, $|\psi_{1}\rangle,|\psi_{2}\rangle$ being pure Gaussian states.
\paragraph*{Proof}
It is sufficient to consider mixtures comprising a finite number of pure Gaussian states. Infinite sums and integrals [such as that appearing in the definition \eqref{setG}] are included in the discussion via a limiting procedure, thanks to the continuity of $\Phi$. Suppose $B(n)\!=\!\Phi(\rho_{n})$, $\rho_{n}\!=\!\!\sum_j p_j|\psi_j\rangle\langle \psi_j|$ and the $|\psi_j\rangle$'s are pure Gaussian states of average photon number $n_j$, such that $\bar n\equiv \sum p_j n_j\leq n$. From Eq.~\eqref{abstract-bound} and the fact that ${\cal G}_n\subset{\cal G}_m$ for $m>n$, it follows that $B(n)$ is a non-increasing function of $n$ (that is, the minimum is in general lower on a larger set). Then,
\begin{align}
	&B(\bar n)\geq B(n)=\Phi[\rho_{n}]=\Phi[\sum_j p_j|\psi_j\rangle\langle \psi_j|]\nonumber\\
	&=\sum_j p_j\Phi[|\psi_j\rangle\langle \psi_j|]\geq \sum_j p_j B(n_j)\geq B(\bar n),\label{concave}
\end{align}
where the first inequality follows from the non-increasing behaviour of $B$, the second inequality follows from the fact that $|\psi_j\rangle\in{\cal G}_{n_j}$ may not be the state minimizing $\Phi$ on ${\cal G}_{n_j}$, and the third from the convexity of $B$ proven in Lemma 1. To avoid contradiction, only the equal signs are possible in Eq.~\eqref{concave}.
This also implies that for all pure states $|\psi_j\rangle$ involved in the sum it must be 
\begin{equation}
\Phi[|\psi_j\rangle\langle \psi_j|]=B(n_j).\label{pure-tight}
\end{equation}
If $n_J=n$ for some $J$, then $B(n)=\Phi[|\psi_{J}\rangle\langle \psi_{J}|]$.\\
Otherwise, there must be at least two values $j_1$ and $j_2$ in the sum such that $n_{j_1}\!<\!\bar n\!<\!n_{j_2}$, thus one can find $0\!\!<\!\!q\!\!<\!\!1$ yielding $q n_{j_1}\!+\!(1\!-\!q)n_{j_2}\!=\!\bar n$. For $r$ sufficiently small (and positive), it is possible to decompose 
\begin{align}
	\sum p_j\proj{\psi_{j}}&=r[q \proj{\psi_{j_1}}\!+\!(1\!-\!q)\proj{\psi_{j_2}}]\nonumber\\
	&+(1-r)\sum\tilde p_j\proj{\psi_{j}}
\end{align}
where $\{\tilde p_j\}$ is a probability distribution such that $\sum\tilde p_jn_j=\bar n$. Correspondingly, Eq.~\eqref{concave} implies
\begin{align}
	B(\bar n)&=r\left(q \Phi[\proj{\psi_{j_1}}]\!+\!(1\!-\!q)\Phi[\proj{\psi_{j_2}}]\right)\nonumber\\
	&+(1-r)\sum\tilde p_j\Phi[\proj{\psi_{j}}]\nonumber\\
	&=r [qB(n_{j_1})\!+\!(1\!-\!q)B(n_{j_2})]+(1-r)\sum \tilde p_j B(n_j)\nonumber\\
	&\geq r B(\bar n)+(1\!-\!r)B(\bar n)=B(\bar n),
\end{align}
where we have exploited Eq.~\eqref{pure-tight} and the convexity of $B$. This implies that it must be $qB(n_{j_1})\!+\!(1\!-\!q)B(n_{j_2})=B(\bar n)$. Moreover, we had $B(n)=B(\bar n)$. Therefore
\begin{equation}
	B(n)=q\Phi[\proj{\psi_{j_1}}]\!+\!(1\!-\!q)\Phi[\proj{\psi_{j_2}}].
\end{equation}
We have thus proven that $B(n)$ is either achieved by a pure Gaussian state or by a Rank-2 mixture of pure Gaussian states.
\section{Properties of the quasiprobability bounds}\label{quasi-general}
For a single-mode pure Gaussian quantum state $|\psi_G\rangle=D(\alpha)S(\xi)|0\rangle$, with $\alpha=|\alpha|{\rm e}^{i\theta},\xi=r {\rm e}^{i\phi}$ the value of the $s$-parametrized quasiprobability distribution in the origin can be written as 
\begin{equation}
\label{eq:Quasi}
Q_s[|\psi_G\rangle\!\langle\psi_G|](0) = \frac{2e^{-\frac{2(n-m)(1+2m-2\sqrt{m(1+m)}\cos(2\theta-\phi)-s)}{1+s(s-2-4m)}}}{\pi\sqrt{1+s(s-2-4m)}}
\end{equation}
where $n=|\alpha|^2 + \sinh^2 r$ is the average number of photons and $m = \sinh^2 r\leq n$ is the squeezing fraction. The condition $s<0$ ensures that the expression in Eq.~\eqref{eq:Quasi} is real. 
From this expression, we can prove some general properties of the functions $B_s(n)$. Firstly, we notice that $B_s(n)>0$ for any $n\geq0$. Also, since the only state in ${\cal G}_0$ is the vacuum $|0\rangle$, we have
\begin{equation}
B_s(0)=\frac{2}{\pi\sqrt{1+s(s-2)}}.
\end{equation}
One can also see that $\lim_{n\to\infty}B_s(n)=0$ for any $s<0$. Then, it is easy to show that $B_s(n)$ is strictly decreasing: suppose that $\widetilde{n}>n$ but $B_s(\widetilde{n})=B_s(n)$. Since $B_s$ tends to zero for large $n$, it is possible to find $N>\widetilde{n}>n$ such that $B_s(N)<B_s(\widetilde{n})$, and $q\in(0,1)$ such that $qn+(1-q)N=\widetilde{n}$. Then one would obtain $qB_s(n)+(1-q)B_s(N)\geq B_s(\widetilde{n})=B_s(n)$, on the other hand $qB_s(n)+(1-q)B_s(N)$ $<qB_s(n)+(1-q)B_s(\widetilde{n})$ $=B_s(n)$ thus reaching a contradiction.

Finally, we show  that the bound $B_s(n)$ is achieved by a state with $n$ average photons, that is $B_s(n)=Q_s[\rho_n](0)$ and ${\sf Tr}[\rho_n a^\dag a]=n$ (for brevity, in what follows we shall omit the phase space argument of $Q_s$, assuming it to be always ``$(0)$''). Assuming that this is not the case, we write $B_s(n)=Q_s[\rho_{\widetilde{n}}]$, s.t. ${\sf Tr}[\rho_{\widetilde{n}} a^\dag a] =\widetilde{n}<n$. However, one has $\rho_{\widetilde{n}} \in \mathcal{G}_{\widetilde{n}}\subset \mathcal{G}_n$, and as a consequence we reach the absurd conclusion $B_s(n)=Q_s[\rho_{\widetilde{n}}]\geq B_s(\widetilde{n})$, which is in contradiction with the strict monotonicity of $B_s$.  \\

We remark that all the properties derived in this section hold for any linear functional whose bound satisfies the properties: (i) $B(0) > 0$ and (ii) $\lim_{n\to\infty}B(n)=0$.
\section{Near-Optimality of pure states}\label{pure-quasi}
We note that in general it must be $B^P_s(n)\geq B_s(n)$, since we can not exclude that the minimum may be reached by a Rank-2 state. By adopting the same reasoning as in Ref.~\cite{QNGWigner}, however, one can show that if $B^P_s(n)$ is convex in the variable $n$, then it must be that $B^P_s(n)=B_s(n)$. While we have the conjecture that this is the case for any $s\leq0$, the functional form of $B^P_s$ is in general too cumbersome to verify its convexity analytically. Adopting a numerical approach we have verified that, for the values $s=\{ -1/4, -1/2, -1,-2,-3\}$, the function $B^P_s(n)$ is convex for $n\leq n_{\sf max}\simeq 10^{15}$ (that is, its second derivative is positive) \cite{numerics}. 

Then, using the results of Appendix~\ref{genproof} we note that the only possibility to have $B^P_s(n)\neq B_s(n)$ is when $B_s(n)=(1-p)B_s^P(n_1)+pB_s^P(n_2)$, with the average photon numbers of the two pure Gaussian states and the probability $p$ verifying respectively $n_1<n$, $n_2>n_{\sf max}$, and $p<n/n_{\sf max}$. Thus,  we have
\begin{align}
0&\leq B_s^P(n)-B_s(n)=B^P_s(n)\!-\!B^P_s(n_1)\nonumber\\
&+p\left[B^P_s(n_1)-B^P_s(n_2)\right]\leq p B^P_s(n_1)\nonumber\\
&<\frac{2}{\pi\sqrt{1\!+\!s(s\!-\!2)}}\frac{n}{n_{\sf max}}\sim10^{-15}n,
\end{align}
where we have used $B^P_s(n)\!-\!B^P_s(n_1)\leq0$, which follows from $n_1<n$ and $B_s^P(n)$ being monotonically decreasing in $n$ [this can be seen easily from the abstract definition in Eq~\eqref{purebound}, o more directly from Eq.~\eqref{eq:QsminA}].\\

\section{Pure state bounds}\label{purestatebounds}
Our goal is now to minimize the function in Eq.~\eqref{eq:Quasi} over all pure Gaussian states with average photon number $n$. We first notice that setting $2\theta-\phi=\pi$ yields the inequality
\begin{equation}
Q_s[|\psi_G\rangle\!\langle\psi_G|](0) \geq\frac{2e^{-\frac{2(n-m)(1+2m+2\sqrt{m(1+m)}-s)}{1+s(s-2-4m)}}}{\pi\sqrt{1+s(s-2-4m)}}. \label{eq:QsminA}
\end{equation}
Finally, one has to minimize the above expression with respect to the squeezing fraction $m$, under the constraint $m\leq n$. The optimizing value of $m$ for a given $s$ is denoted $m_s(n)$. As an example, we here consider the case $s=-1$, where the function $Q_{-1}$ 
is the so-called Husimi Q-function
\begin{equation}
Q_{-1}[\varrho] (\alpha) = \frac{\langle \alpha | \varrho | \alpha\rangle}{\pi}.
\end{equation}
Here $|\alpha\rangle= D(\alpha)|0\rangle$ denotes a coherent state, showing that the Husimi Q-function corresponds to the heterodyne probability distribution of the quantum state $\varrho$. For a generic state $\varrho \in \mathcal{G}$, the value of the squeezed photon number for which the Husimi Q-function in the origin 
is minimized is:
\begin{widetext}

\begin{align}
m_{-1}(n) &= \frac{1}{6} \left(2 (n-1) + \sqrt{3}\ \mathrm{Im}[ g(n)]  - \mathrm{Re}[g(n)]\right),\\
\mathrm{where}\ g(n) &= (-17-21n+3n^2+8n^3+3i(1+n)\sqrt{6+3n(24+n(37+16n))})^{1/3}.
\end{align}
%\begin{align}
%m_{-1}(n) &= \frac{1}{3}(n-1)+\frac{i(i+\sqrt{3})(7+2n(5+2n))}{12(-17-21n+3n^2+8n^3+3i(1+n)\sqrt{6+3n(24+n(37+16n))})^{1/3}} \\
%\nonumber&\ \ \ -\frac{1}{12}(1+i\sqrt{3})(-17-21n+3n^2+8n^3+3i(1+n)\sqrt{6+3n(24+n(37+16n))})^{1/3}.
%\end{align}

%\begin{figure}[h!]
%\hfill
%\subfigure[Real part of maximum value of squeezed photon number $m_{-1}(n)$ of the Q function.\label{f:realm}]{\includegraphics[width=0.45\columnwidth]{Realm.pdf}}
%\hfill
%\subfigure[Imaginary part of maximum value of squeezed photon number $m_{-1}(n)$ of the Q function.\label{f:imagm}]{\includegraphics[width=0.45\columnwidth]{Imagm.pdf}}
%\hfill
%\caption{Real and imaginary parts of $m_{-1}(n)$ showing that the negligible imaginary part is a product of numerical noise and is therefore neglected in further calculations.\label{fig:realfn}}
%\end{figure}

The bound $B_{-1}^P(n)$ can be then obtained by substituting this function into the form of Eq. (\ref{eq:QsminA}) where $s=-1$:
\begin{equation}
Q_{-1}[|\psi_G\rangle\!\langle\psi_G|](0) \geq\frac{2e^{-\frac{2(n-m_{{-}1}(n))(1+2m_{{-}1}(n)+2\sqrt{m_{{-}1}(n)(1+m_{{-}1}(n))}+1)}{1+(3+4m_{{-}1}(n))}}}{\pi\sqrt{1+(3+4m_{{-}1}(n))}}. \label{eq:Q1minA}
\end{equation}
\end{widetext}
Where the final result is too cumbersome to be reported here. Identical approaches can be effectively pursued for other values of $s<0$. 

\begin{thebibliography}{99}

\bibitem{Wigner} E. Wigner, Phys. Rev. {\bf 40} 749-759 (1932).
\bibitem{Glauber} R. J. Glauber, Phys. Rev. {\bf 131}, 2766-2788 (1963).
\bibitem{Sudarshan} E. C. G. Sudarshan, Phys. Rev. Lett. {\bf 10} 277-279 (1963).
\bibitem{Lee91} C. T. Lee, Phys. Rev. A {\bf 44}, R2775 (1991).
\bibitem{Lee92} C. T. Lee, Phys. Rev. A {\bf 45}, 6586 (1992).
\bibitem{Arvind97} Arvind, N. Mukunda, R. Simon, Phys. Rev. A {\bf 56}, 5042 (1997).
\bibitem{Arvind98} Arvind, N. Mukunda, R. Simon, J. Phys. A {\bf 31}, 565 (1998).
\bibitem{Marchiolli01} M. A. Marchiolli, V. S. Bagnato, Y. Guimaraes, B. Baseia, Phys. Lett. A {\bf 279}, 294 (2001).
\bibitem{Paris01} M. G. A. Paris, Phys. Lett. A {\bf 289}, 167 (2001).
\bibitem{Dodonov02} V. V. Dodonov, J. Opt. B {\bf 4}, R1 (2002).
\bibitem{Kenfack04} A. Kenfack, K. Zyczkowski, J. Opt. B {\bf 6}, 396 (2004).
\bibitem{Paavola11}
{J. Paavola, M. J. Hall, M. G. A. Paris, S. Maniscalco},
Phys. Rev. A {\bf 84}, 012121 (2011).


\bibitem{Shchukin05} E. V. Shchukin, W. Vogel, Phys. Rev. A {\bf 72}, 043808 (2005).
\bibitem{Kiesel08} T. Kiesel, W. Vogel, V. Parigi, A. Zavatta, M. Bellini, Phys. Rev. A {\bf 78}, 021804 R (2008).
\bibitem{Vogel08} W. Vogel, Phys. Rev. Lett. {\bf 100}, 013605 (2008).
\bibitem{Simon00} R. Simon, Phys. Rev. Lett. {\bf 84}, 2726 (2000).
\bibitem{Duan00} Lu-Ming Duan, G. Giedke, J. I. Cirac, P. Zoller, Phys. Rev. Lett. {\bf 84}, 2722 (2000).
\bibitem{Marian02} P. Marian, T. A. Marian, H. Scutaru, Phys. Rev. Lett. {\bf 88}, 153601 (2002).
\bibitem{Giorda10} P. Giorda, M. G. A. Paris, Phys. Rev. Lett. {\bf 105}, 020503 (2010); G. Adesso, A. Datta, Phys. Rev. Lett. {\bf 105}, 030501 (2010).
\bibitem{Ferraro12} A. Ferraro, M. G. A. Paris, Phys. Rev. Lett. {\bf 108}, 260403 (2012).
\bibitem{Gehrke12} C. Gehrke, J. Sperling, W. Vogel, Phys. Rev. A {\bf 86}, 052118(2012).
\bibitem{Buono10} D. Buono, G. Nocerino, V. D'Auria, A. Porzio, S. Olivares, M. G. A. Paris, J. Opt. Soc. Am. B {\bf 27}, 110 (2010); D. Buono, G. Nocerino, A. Porzio, S. Solimeno. Phys. Rev. A {\bf 86}, 042308 (2012).

\bibitem{Hudson} R. L. Hudson, Rep. Math. Phys. {\bf 6}, 249 (1974).
\bibitem{Werner} T. Brocker and R. F. Werner, J. Math. Phys. {\bf 36}, 62 (1995).
\bibitem{Mandilara} A. Mandilara, E. Karpov and N. J. Cerf, Phys. Rev. A {\bf 79}, 062302 (2009).
\bibitem{mari12} A. Mari and J. Eisert, Phys. Rev. Lett. {\bf 109}, 230503 (2012).
\bibitem{veitch13} V. Veitch, N. Wiebe, C. Ferrie and J. Emerson, New J. Phys. {\bf 15}, 013037 (2013). 

\bibitem{Genoni07} M. G. Genoni, M. G. A. Paris and K. Banaszek, Phys. Rev. A {\bf 76}, 042327 (2007).
\bibitem{Genoni08} M. G. Genoni, M. G. A. Paris and K. Banaszek, Phys. Rev. A {\bf 78}, 060303 (2008).
\bibitem{Genoni10} M. G. Genoni, M. G. A. Paris, Phys. Rev. A {\bf 82}, 052341 (2010).

\bibitem{MarcoPhAdd} M. Barbieri, N. Spagnolo, M. G. Genoni, F. Ferreyrol, R. Blandino, M. G. A. Paris, P. Grangier and R. Tualle-Brouri, Phys. Rev. A {\bf 82}, 063833 (2010).

\bibitem{TWBcond} A. Allevi, A. Andreoni, F. A. Beduini, M. Bondani, M. G. Genoni, S. Olivares and M. G. A. Paris, Eur. Phys. Lett. {\bf 92}, 20007 (2010).
\bibitem{ThermalCond} A. Allevi, A. Andreoni, M. Bondani, M. G. Genoni and S. Olivares, Phys. Rev. A {\bf 82}, 013816 (2010).

\bibitem{BarRad} S. M. Barnett and P. M. Radmore, \textit{Methods in Theoretical Quantum Optics}, Clarendon Press, Oxford (1997).

\bibitem{QNGRadim} R. Filip and L. Mista, Phys. Rev. Lett. {\bf 106}, 200401 (2011).
\bibitem{QNGWigner} M. G. Genoni, M. L. Palma, T. Tufarelli, S. Olivares, M. S. Kim and M. G. A. Paris, Phys. Rev. A  {\bf 87}, 062104 (2013).

\bibitem{Jez11} M. Je\v{z}ek, I. Straka, M. Mi\v{c}uda, M. Du\v{s}ek, J.  Fiura\v{s}ek, 
R. Filip, Phys. Rev. Lett. {\bf 107}, 213602 (2011).
\bibitem{Jez12} M. Je\v{z}ek, A. Tipsmark, R. Dong, J. Fiura\v{s}ek, L.
Mi\v{s}ta, Jr., R. Filip, U. L. Andersen, Phys. Rev. A {\bf 86}, 043813 (2012).
\bibitem{predojevic12}A. Predojevic, M. Jezek, T. Huber, H. Jayakumar, T. Kauten, G. S. Solomon, R. Filip and G. Weihs, Opt. Expr. {\bf 22},
4789 (2014).
\bibitem{MSKTutorial} M. S. Kim, J. Phys. B: At. Mol. Opt. Phys. {\bf 41}, 133001 (2008).

%\bibitem{gaussian} S. D. Bartlett, B. Sanders, S. L. Braunstein, K. Nemoto, Phys. Rev. Lett. {\bf 88}, 097904/1-4 (2002).
\bibitem{convolution} 
K. E. Cahill and R. J. Glauber, Phys Rev. {\bf 177}, 1882 (1969), {\em
ibidem} {\bf 177}, 1857 (1969).
\bibitem{Husimi} Kodi Husimi, Proc. of the Phys.-Math. Soc. of Japan {\bf 22} 264-314 (1940).
\bibitem{QNGCats} M. L. Palma, J. Stammers, M. G. Genoni, T. Tufarelli, S. Olivares, M. S. Kim and M. G. A. Paris, Phys. Scr. {\bf T160}, 014035 (2014)
\bibitem{matteoQSM} M. G. A. Paris, Phys. Rev. A {\bf 53}, 2658 (1996).

\bibitem{numerics} When $n\gtrsim3\cdot 10^{15}$, our numerics become unstable and no conclusive results have been obtained.
\end{thebibliography}
\end{document}